\documentclass[a4paper,11pt]{article}

\usepackage{graphicx}
\usepackage{amsmath}
\usepackage[margin=3.5cm]{geometry}
\providecommand{\FullConference}[1]{\def\theconference{#1}}
\begin{document}

\FullConference{21st International Conference on B-Physics at Frontier Machines (BEAUTY 2026)\\
1 to 5 June, 2026\\
Maastricht, the Netherlands}

\title{New measurement of BR$(K^+\rightarrow\pi^+\nu\bar\nu)$ at the NA62 experiment}

\author{Silvia Martellotti
\thanks{silvia.martellotti@lnf.infn.it}\textsuperscript{,\,$\dagger$}
\\[1ex] \small INFN Laboratori Nazionali di Frascati\\ \small Via Enrico Fermi, 54, 00044 Frascati (Rome), Italy.
\\[2ex] \footnotesize\itshape \begin{tabular}{c}\theconference\end{tabular}
}
\date{}
\maketitle
\begin{abstract}
The decay $K^{+} \rightarrow \pi^{+}\nu\overline{\nu}$ with a very precisely predicted branching ratio of less than $10^{-10}$ is one of the best
candidates to reveal indirect effects of new physics at the highest mass scales. 
The NA62 experiment at the CERN SPS is designed to measure the branching ratio of the $K^{+} \rightarrow \pi^{+}\nu\overline{\nu}$ decay. 
A new measurement of its branching ratio based on the 2023–2024 dataset is presented. 
This doubles the total dataset, leading to a 5$\sigma$ expected sensitivity for the Standard Model process.
\end{abstract}
\dimen\footins=0.45\textheight
\interfootnotelinepenalty=0
{\renewcommand{\thefootnote}{$\dagger$}%
\footnotetext{for the NA62 Collaboration: A.~T.~Akmete, R.~Aliberti, S.~Alibocus, F.~Ambrosino, R.~Ammendola, B.~Angelucci, A.~Antonelli, G.~Anzivino, R.~Arcidiacono, T.~Bache
% ...
}}

\section{The $K^+ \rightarrow \pi^+ \nu\overline{\nu}$ decay in the Standard Model}
The $K^+\rightarrow\pi^+\nu\overline{\nu}\, (K_{\pi\nu\overline{\nu}})$
decay is a flavour changing neutral current process proceeding through
box and electroweak penguin diagrams. The process is very rare, due to
quadratic GIM mechanism and strong Cabibbo suppression. The dominant
contribution comes from the short-distance physics of the top quark
loop, with a small charm quark contribution and long-distance
corrections. This makes $K_{\pi\nu\overline{\nu}}$
theoretically clean: the Standard Model (SM) predictions, using CKM
matrix elements extracted from tree-level processes, give
$\mathrm{BR}(K^+ \rightarrow \pi^+\nu\overline{\nu}) \approx 8 \times 10^{-11}$ with a precision better than 8\% \cite{ref:sm_1,ref:sm_2}.
The uncertainties on the predictions are dominated by the external inputs \cite{ref:loop, ref:precision}. An experimental measurement of this branching ratio is, therefore, a powerful test of the SM. 
Furthermore, such a precise and small branching ratio makes $K^+ \rightarrow \pi^+ \nu\overline{\nu}$ decay
extremely sensitive to various BSM effects, probing new physics at mass
scales up to $\mathcal{O}(100)$~TeV~\cite{ref:np1}. In several BSM scenarios large modifications to the branching ratio
are predicted, as well as correlations with other flavour observables \cite{ref:np2} and the neutral kaon decay
mode $K_L \rightarrow \pi^0 \nu\overline{\nu}$ \cite{ref:np3}.

\section{NA62 beam and detector}

With the main goal of measuring the branching ratio (BR) of the ultra-rare $K^+\rightarrow\pi^+\nu\bar\nu$ decay, the NA62 experiment started physics data taking in 2016 and has collected the world’s largest data sample of $K^+$ decays. 

\begin{figure}[htbp]
\includegraphics[width=14cm]{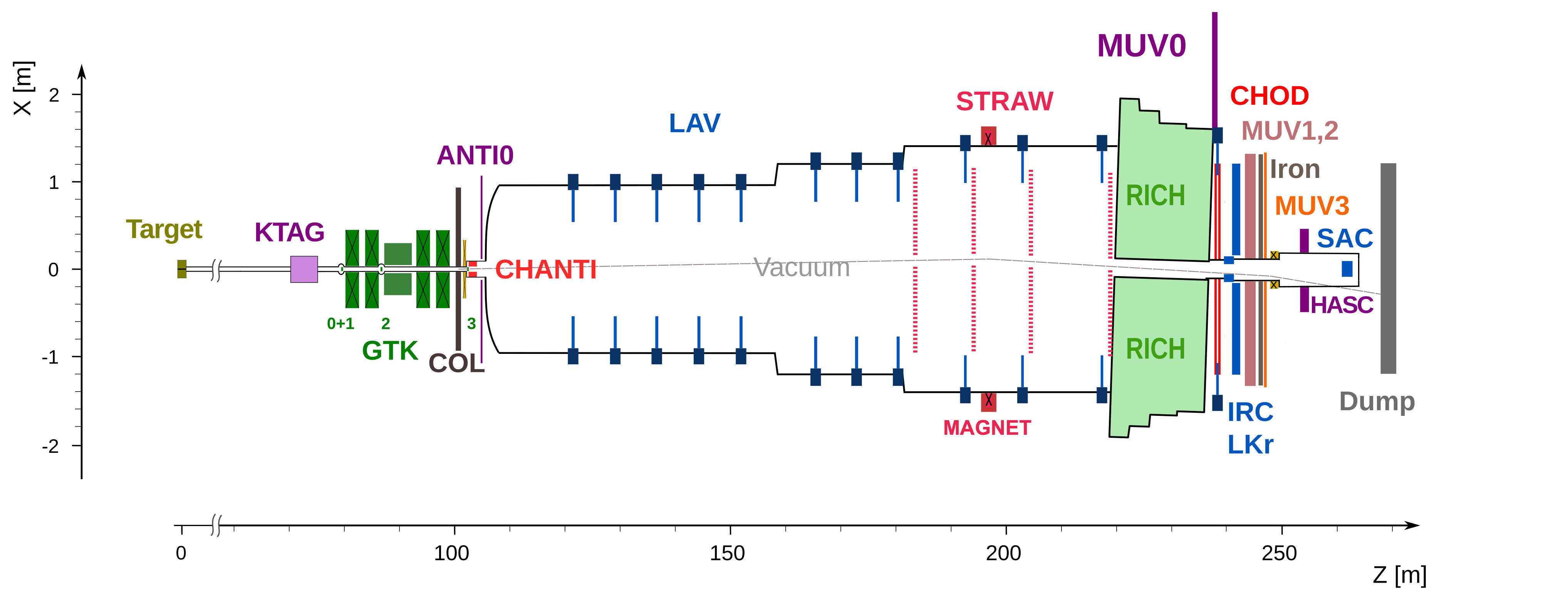} 
\caption{Schematic view of the NA62 detector layout in the XZ plane from 2021 onwards.} 
\label{fig:setup}
\end{figure}
The NA62 experiment is located at the CERN SPS and operates with a 75~GeV$/c$ hadron beam with a 6\% $K^+$ component. 
A schematic view of the NA62 detector layout in the XZ plane is shown in Fig.~\ref{fig:setup}. An extensive description of the beam line and detector systems can be found in~\cite{ref:na62_1}.
In parallel to the $K^+\rightarrow\pi^+\nu\bar\nu$ trigger, additional trigger lines select other event topologies~\cite{ref:trigger}. 
The incoming kaons are identified by a differential Cherenkov detector (KTAG) with $\sigma \sim 70$ ps time resolution and the kaon momenta are measured by a four-station silicon pixel beam spectrometer (GTK) before they enter a vacuum tank containing a 65 m long fiducial decay volume (FV). The Veto Counter (VC), ANTI0, and CHANTI detectors reject the decays and interactions of beam particles occurring between the first and last GTK stations. Downstream of the FV, the momenta of charged particles are measured by a magnetic spectrometer (STRAW). A Ring Imaging Cherenkov detector (RICH) provides a trigger time with $\sigma \sim 70$ ps time resolution and is optimized for the identification of positively charged particles. The presence of signals in the quadrants of a Charged Particle Hodoscope (CHOD) is also used as a trigger condition. Additional $\pi/\mu$ discrimination is provided by the quasi-homogeneous liquid krypton electromagnetic calorimeter (LKr) and hadronic sampling calorimeters (MUV1,2), together with an array of scintillator tiles (MUV3) located behind an 80 cm iron wall, used to provide a muon veto in the trigger.  A comprehensive set of photon veto detectors provides hermetic coverage of polar angles up to 50 mrad from the beam axis. This comprises twelve large-angle veto (LAV) stations composed of lead-glass blocks, the LKr, and the intermediate-ring (IRC) and small-angle (SAC) lead/scintillator shashlik calorimeters.

\section{The BR($K^{+} \rightarrow \pi^{+}\nu\overline{\nu}$) measurement strategy}

The analysis of the 2023–2024 dataset is presented here~\cite{ref:pnn}.
The $K_{\pi\nu\overline{\nu}}$ decay signature is one track in the initial and final state and nothing else, as the neutrinos escape detection. The main kinematic variable is $m^2_{\mathrm{miss}} \equiv\left(p_K - p_{\pi ^+} \right)^2$, where $p_K$ and $p_{\pi^+}$ are the 4-momenta of the $K^+$ and $\pi^+$ respectively. 
Two signal regions are chosen (R1 and R2 shown in Fig.~\ref{fig:result} (left) and Fig.~\ref{fig:summary} (left)) using $m^2_{\mathrm{miss}}$ and the $\pi^+$ momentum, to maximise access to the signal phase-space while
avoiding the 3 main $K^+$ backgrounds: $K^+ \rightarrow \mu^+ \nu_\mu$, $K^+ \rightarrow \pi^+ \pi^0$ and $K^+ \rightarrow \pi^+ \pi^+ \pi^-$.

Events with a single downstream track are selected using the STRAW, CHOD and RICH detectors, with a time resolution of about 100~ps. The track is associated with an in-time kaon identified by the KTAG and with a beam track reconstructed by the GTK, which measures time and momentum of all beam particles. The kaon decay vertex, defined at the point of closest approach between the GTK and STRAW tracks, must lie inside the 65~m long FV.
The $\pi^+$ is identified by the RICH detector, 
additional $\pi/\mu$ discrimination is provided by LKr and MUV calorimeters,
providing a $10^7$ $\mu^+$ suppression for a 75\% $\pi^+$
identification efficiency. Further background rejection is provided by the veto 
detectors (LAVs, LKr, IRC, SAC). The $\pi^0$ rejection inefficiency, 
defined as the probability that a $K^+ \rightarrow \pi^+ \pi^0$ candidate
passing the normalisation selection is not rejected by the photon or multiplicity veto criteria, is measured to be $(1.24 \pm 0.04)\times 10^{-8}$, when restricting the $\pi^+$ momenta to the 15--45~GeV$/c$ range. This is 30\% better than for 2021–2022 data
due to improved data quality control and optimised beam intensity.

\subsection{Improvements in the 2023–2024 data taking and analysis}

Based on the study of the signal yield per spill as a function of the
secondary beam intensity, since August 2023 NA62 has operated at the optimal intensity,
around 75\% of the design value~\cite{ref:spsc}. This choice maximises the readout efficiency and reduces the random veto losses.
Hardware upgrades were also introduced in 2023: the Cherenkov detector for Kaon identification (KTAG) has been replaced with one optimized for and filled with Hydrogen, instead of nitrogen, to reduce the
material budget~\cite{ref:ktag}, and the readout of the VC and CHANTI detectors was
upgraded with a prototype FPGA-based TDC system read out by an ATLAS FELIX board.
Both the GTK beam track reconstruction and the calorimetric particle identification
were also improved, exploiting Machine Learning algorithms.

\section{Results}

\begin{figure}[htbp]
\noindent\hspace*{\fill}%
\includegraphics[height=4.4cm]{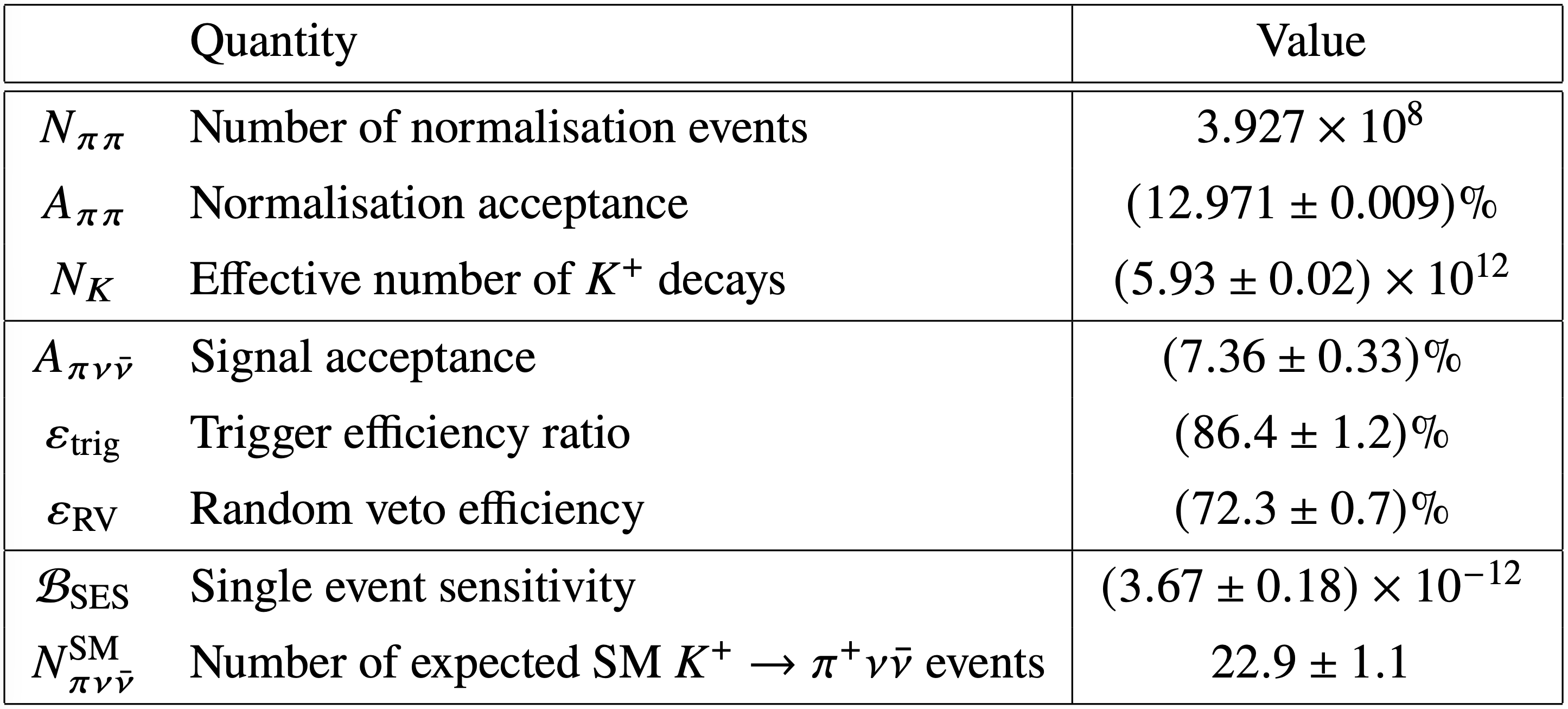}%
\hspace*{\fill}%
\includegraphics[height=4.4cm]{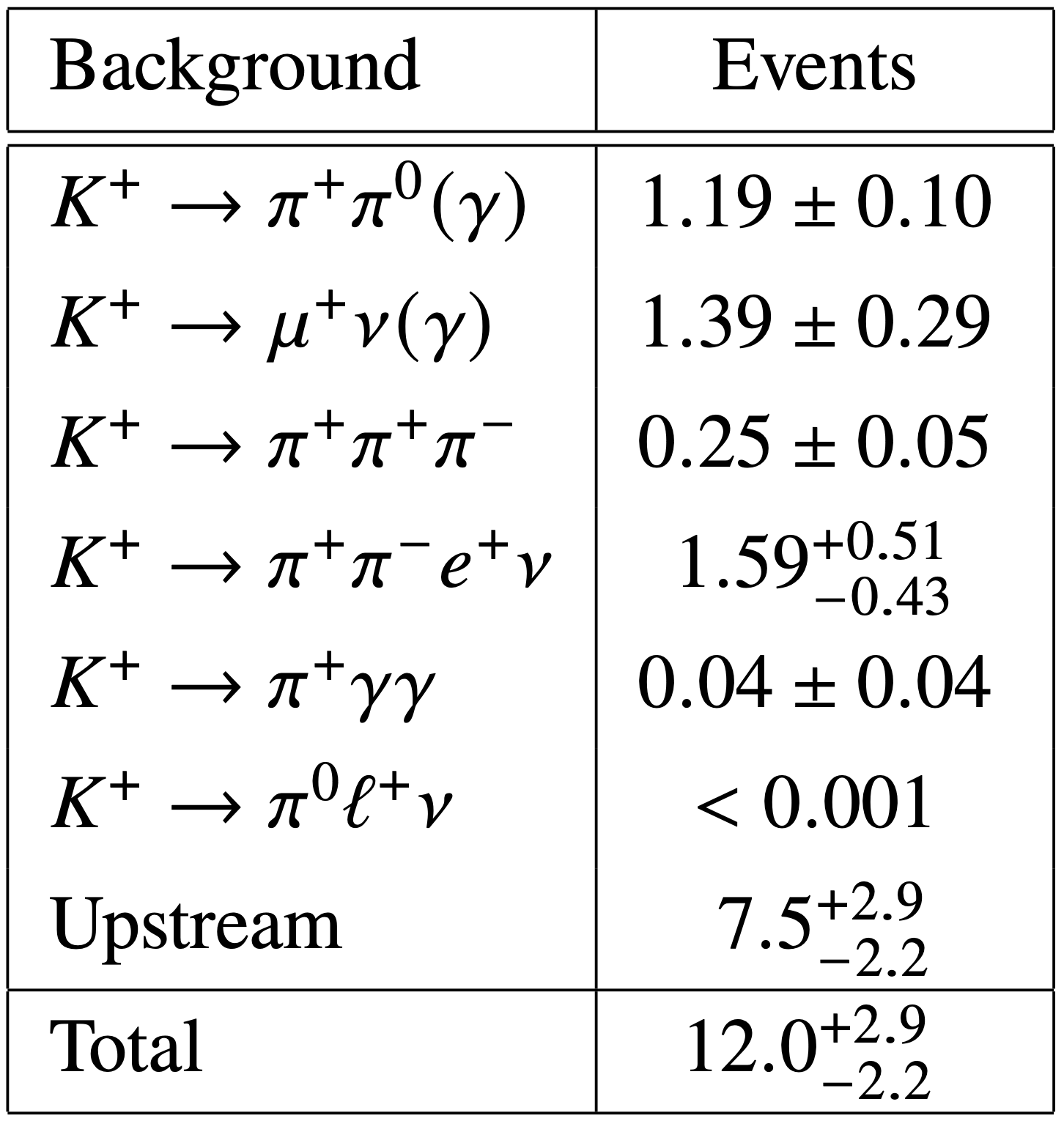}%
\hspace*{\fill}%
\caption{Left table: inputs to the signal sensitivity for the 2023–2024 data sample. Right table: background estimates in the two signal regions for
2023–2024 data, summed over the six $\pi^+$ momentum bins.}
\label{fig:table}
\end{figure}

The effective number of $K^+$ decays of the analysed sample is evaluated as
$N_K = (N_{\pi\pi} \cdot D)/(\mathrm{BR}_{\pi\pi} \cdot A_{\pi\pi})$,
where $N_{\pi\pi}$ is the number of normalisation $K^{+} \rightarrow \pi^{+}\pi^0$ events selected, $D = 400$ is the downscaling factor of the normalisation trigger line, $\mathrm{BR}_{\pi\pi} = \mathrm{BR}(K^{+} \rightarrow \pi^{+}\pi^0) = (20.67 \pm 0.08)\%$, and $A_{\pi\pi}$ is the normalisation selection acceptance. The relative background contamination of the normalisation sample is less than $10^{-4}$.
The expected number of SM $K^{+} \rightarrow \pi^{+}\nu\overline{\nu}$
events is calculated, in 5~GeV$/c$ wide momentum bins ($p_i$), as:
\begin{equation}
N_{\pi\nu\bar\nu}^{\mathrm{SM}}(p_i) = \frac{\mathrm{BR}_{\pi\nu\bar\nu}^{\mathrm{SM}}}{\mathrm{BR}_{\mathrm{SES}}(p_i)} = \mathrm{BR}_{\pi\nu\bar\nu}^{\mathrm{SM}} \cdot N_K(p_i)\cdot A_{\pi\nu\bar\nu}(p_i)\cdot \epsilon_{\mathrm{trig}}(p_i)\cdot \epsilon_{\mathrm{RV}},
\end{equation}
where $A_{\pi\nu\bar\nu}$ is the signal selection acceptance, $\epsilon_{\mathrm{trig}}$ is the ratio of trigger efficiencies for the $\pi\nu\bar\nu$ and normalisation trigger lines for the signal and $K_{\pi^+\pi^0}$ selections respectively,
and $1-\epsilon_{\mathrm{RV}}$ is the probability that a signal event is rejected by the photon and multiplicity veto conditions due to the presence of accidental activity. The latter is independent of
$P_{\pi^+}$ and only depends on the beam intensity. These quantities and their uncertainties are
listed in the left table of Fig.~\ref{fig:table}.
The expected number of SM signal events is $N_{\pi\nu\bar\nu}^{\mathrm{SM}} = 22.9 \pm 1.1$~\cite{ref:pnn}, a factor of 2.3 larger than the expectation from the 2021–2022 data analysis.

The background estimates are presented in the right table of Fig.~\ref{fig:table}.
A data-driven method is used to evaluate backgrounds from the $K^+ \rightarrow \pi^+ \pi^0$, $K^+ \rightarrow \mu^+ \nu$ 
and $K^+ \rightarrow \pi^+ \pi^+\pi^-$ decays, which arise primarily due to misreconstruction of the $m^2_{\mathrm{miss}}$. 
The $K^+ \rightarrow \pi^+ \pi^- e^+\nu$ background is evaluated using $2.5\times 10^9$
simulated events. Other backgrounds from $K^+$ decays in the FV are found to be negligible. The
upstream background, consisting of $K^+$ decays upstream of the FV and inelastic beam-detector interactions between the first and last GTK stations, is evaluated with an updated, fully data-driven strategy~\cite{ref:pnn}. The total expected background is found to be $12.0^{+2.9}_{-2.2}$.

\begin{figure}[htbp]
\noindent\hspace*{\fill}%
\includegraphics[height=5.0cm]{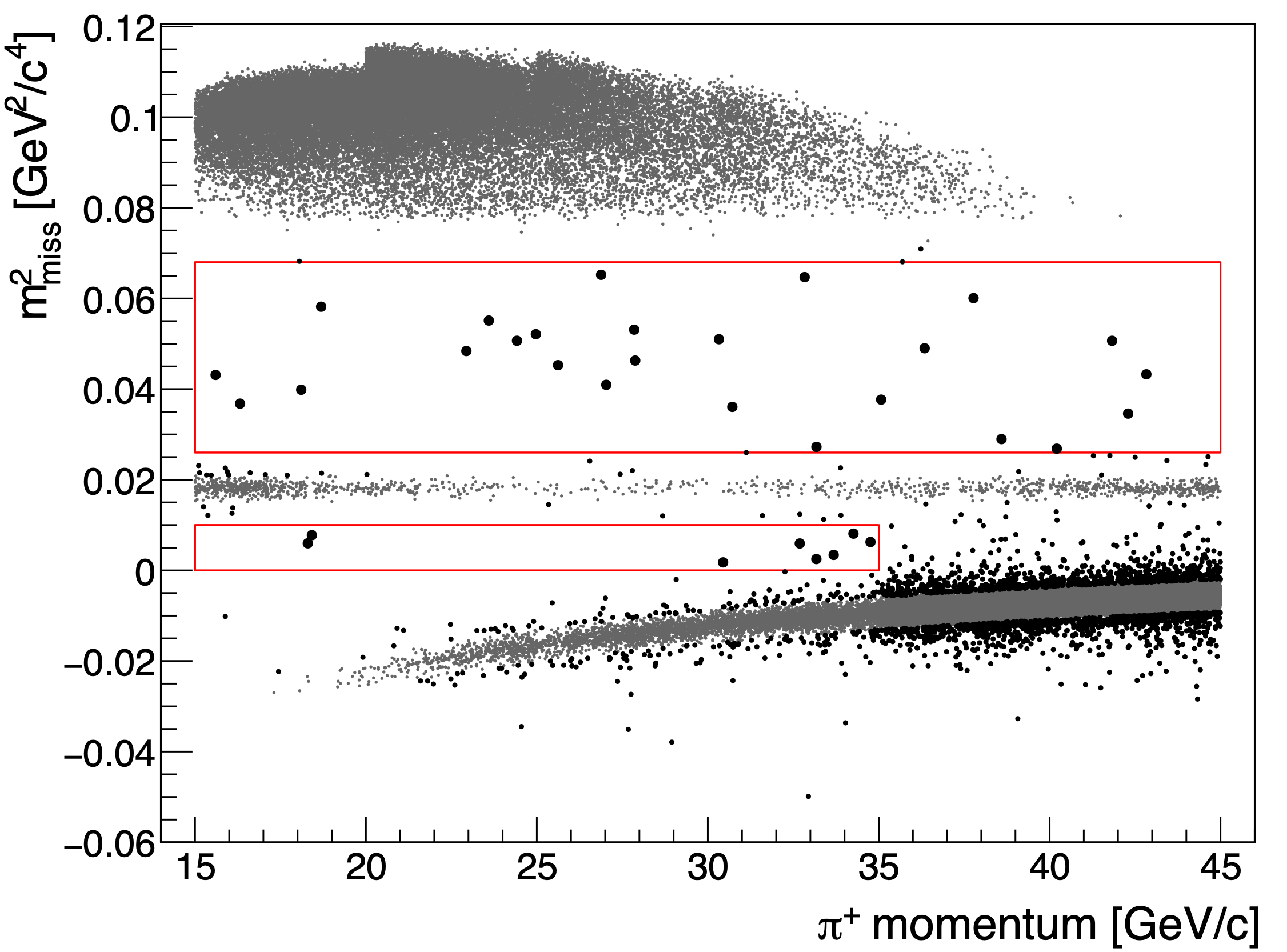}%
\hspace*{\fill}%
\includegraphics[height=5.1cm]{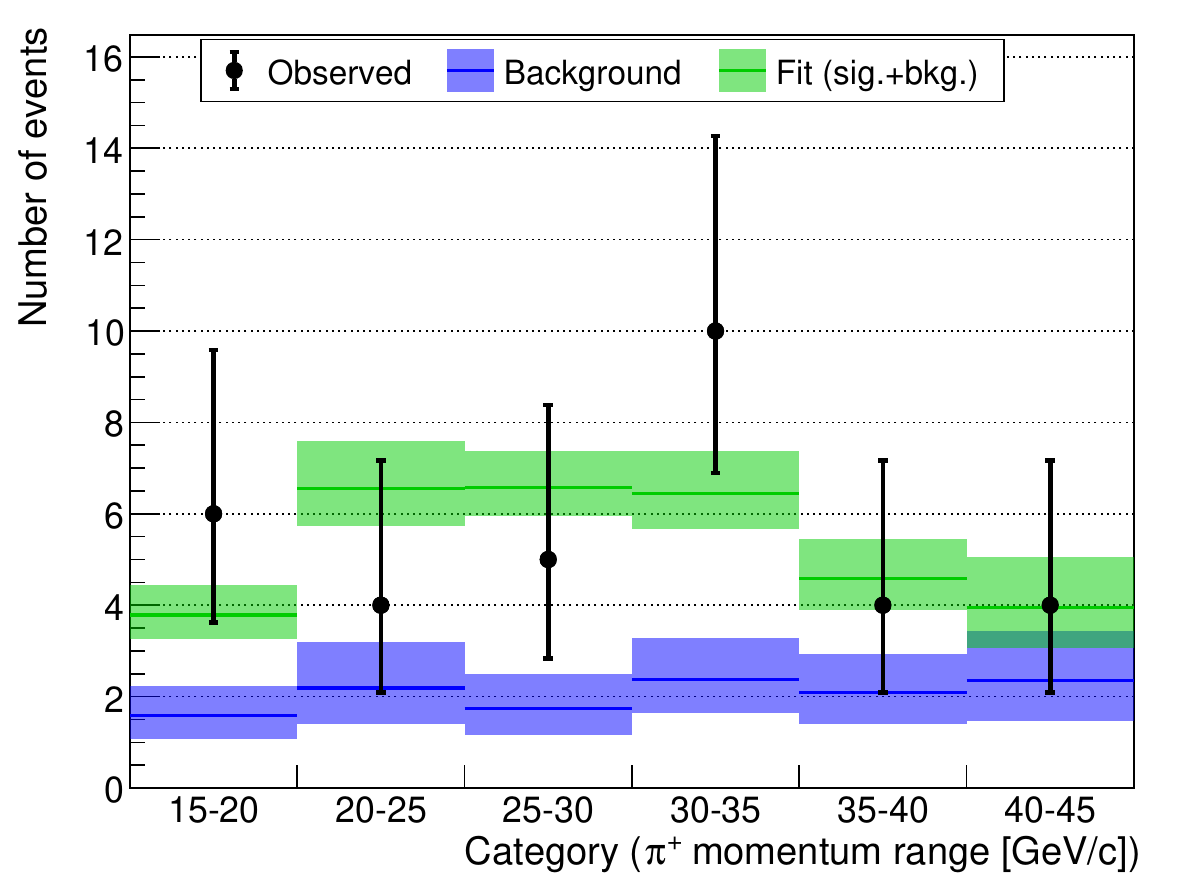}%
\hspace*{\fill}%
\caption{Left: distribution in the $(P_{\pi^+}, m^2_{\mathrm{miss}})$ plane of the observed events in the 2023–2024 dataset
satisfying the signal selection criteria. Red boxes show the signal regions. Right: numbers of expected and observed events 
in the six categories used for the fit of 2023–2024 data.}
\label{fig:result}
\end{figure}

\begin{figure}[htbp]
\noindent\hspace*{\fill}%
\includegraphics[height=5.0cm]{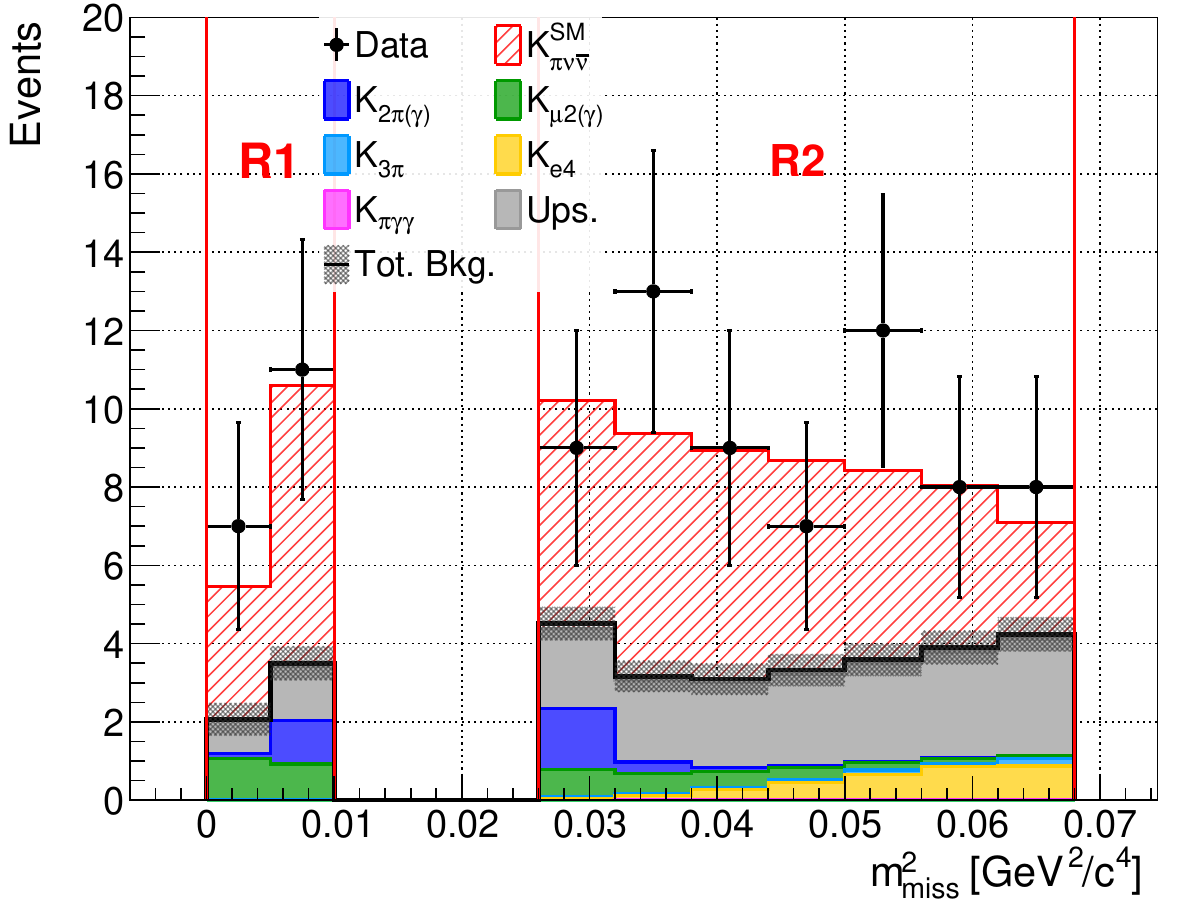}%
\hspace*{\fill}%
\includegraphics[height=5.0cm]{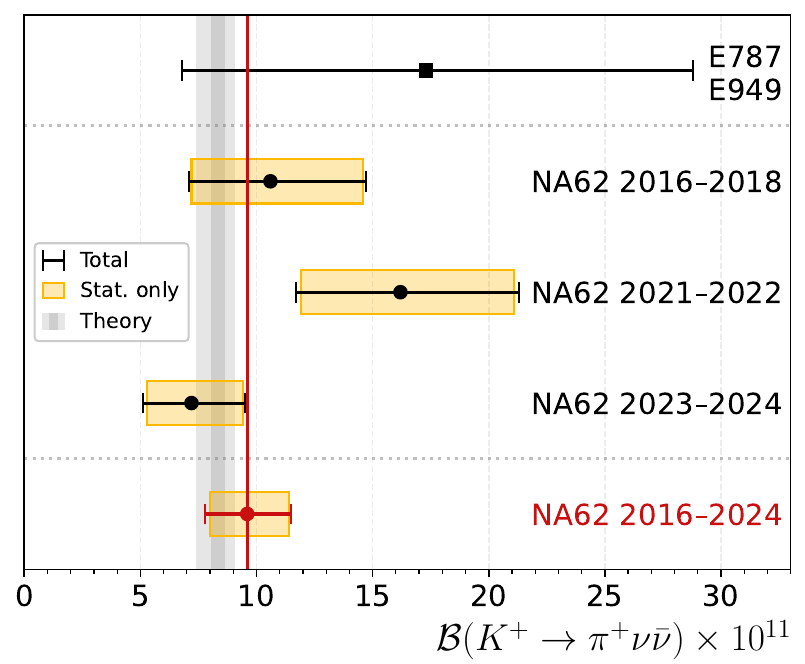}%
\hspace*{\fill}%
\caption{Left: distribution in $m^2_{\mathrm{miss}}$ of the observed events in the 2016–2024 dataset, with SM signal
expectation and estimated backgrounds. R1 and R2 denote the two signal regions. Right: summary of $K^{+} \rightarrow \pi^{+}\nu\overline{\nu}$ 
branching ratio measurements, compared to the SM prediction.}
\label{fig:summary}
\end{figure}

Using the 2016–2024 dataset, including the previously published results, a total expected
background of $30^{+4}_{-3}$ and a total signal expectation of $43\pm 1$ are found. This
corresponds to an expected significance for the SM $K^{+} \rightarrow \pi^{+}\nu\overline{\nu}$
in excess of $5\sigma$ for the first time. In the signal regions 33 signal candidates are found in the 
2023–2024 dataset. The distribution of these events in the $(P_{\pi^+}, m^2_{\mathrm{miss}})$ plane is shown in 
Fig.~\ref{fig:result} (left). Using a profile likelihood-ratio test statistic in a fit with six categories,
corresponding to the momentum bins (Fig.~\ref{fig:result} (right)), the branching ratio is measured to be:
\begin{equation}
\mathrm{BR}_{\text{2023--2024}} (K^{+} \rightarrow \pi^{+}\nu\overline{\nu})= 7.2^{+2.2}_{-1.9}|_{\mathrm{stat}}
\pm 0.9 |_{\mathrm{syst}} \times 10^{-11} = 7.2^{+2.3}_{-2.1} \times 10^{-11}.
\end{equation}
Combining the six 2023–2024 categories with the fifteen 2016–2022 ones~\cite{ref:prev},
the BR is measured to be:
\begin{equation}
\mathrm{BR}_{\text{2016--2024}} (K^{+} \rightarrow \pi^{+}\nu\overline{\nu})= 9.6^{+1.8}_{-1.6}|_{\mathrm{stat}}
{}^{+0.8}_{-0.6}|_{\mathrm{syst}} \times 10^{-11} = 9.6^{+1.9}_{-1.8} \times 10^{-11}.
\end{equation}
A total of 84 candidate events have been observed in the 2016–2024 data, and the distribution of
these events as a function of $m^2_{\mathrm{miss}}$ is shown in Fig.~\ref{fig:summary} (left) together with the evolution of the NA62 results (right).

\section{Conclusions}
Using 2016–2024 data, NA62 has measured the branching ratio of one of the rarest decays ever observed:
$\mathrm{BR} (K^{+} \rightarrow \pi^{+}\nu\overline{\nu}) = 9.6^{+1.9}_{-1.8} \times 10^{-11}$, compatible with the SM predictions.
The inclusion of the dataset collected in 2023–2024 provides a major improvement in precision over the measurement using 2016–2022
data, doubling the effective sample size while significantly decreasing the relative background, with an overall relative uncertainty of 20\%.
2026 is the last year of NA62 data taking; the analysis of the full dataset is ongoing and is expected to improve the
precision to 15\%.



\end{document}